\documentclass[reprint,9pt]{revtex4-2}

\usepackage[utf8]{inputenc}
\usepackage{caption}
\usepackage{color}
\usepackage{dcolumn}
\usepackage{bm}
\usepackage{url}
\usepackage{amsmath}
\usepackage{amssymb}
\usepackage{amsfonts}

\usepackage{xcolor}

\usepackage{graphicx}

\usepackage{listings}

\usepackage{booktabs}
\usepackage{array}
\newcommand{\yes}{\ensuremath{\checkmark}}
\newcommand{\no}{\ensuremath{\times}}
\newcolumntype{P}[1]{>{\centering\arraybackslash}p{#1}}

\begin{document}

\title{Sustainable Packaging of Quantum Chemistry Software with the Nix Package Manager}

\author{Markus Kowalewski}
\email{markus.kowalewski@fysik.su.se}
\affiliation{Department of Physics, Stockholm University, Sweden}

\author{Phillip Seeber}
\email{phillip.seeber@uni-jena.de}
\affiliation{Institute of Physical Chemistry, Friedrich Schiller University Jena, Germany}

\begin{abstract}
The installation of quantum chemistry software packages is commonly done manually and can be
a time-consuming and complicated process. An update of the underlying Linux system
requires a reinstallation in many cases and can quietly break software installed on the system.
In this paper, we present an approach that allows for an easy installation of quantum chemistry
software packages, which is also independent of operating system updates. The use of the Nix package manager allows building software in a reproducible manner, which allows for a
reconstruction of the software for later reproduction of scientific results. The build recipes
that are provided can be readily used by anyone to avoid complex installation procedures.
\end{abstract}

\date{\today}%
\maketitle

\makeatletter
  \renewcommand\@biblabel[1]{#1.}
\makeatother

\renewcommand{\baselinestretch}{1.5}
\normalsize

\renewcommand{\arraystretch}{0.75}

\clearpage

\section{Introduction}
Open-source quantum chemistry program packages are usually compiled manually on a work station,  single compute node, or a high-performance computing system.
This process can be time-consuming and complex, especially
when it has to be carried out for multiple programs. Such a manual installation is in general hard
to replicate, unless its preparation and the use of configuration flags has been meticulously documented. A major problem with such an approach is that the program package will
depend on operating system libraries, unless it has been linked completely statically.
As a consequence, an update of the operating system or any other dependency may quietly break
the software package and eventually make a rebuild necessary.

Another issue that arises from such a manual build strategy is the fact that scientific results
can not be exactly reproduced. To rebuild an old version in exactly the same way, one
needs also the exact states of all dependencies. This problem can in principle be solved by
container solutions \cite{containers}, such as docker \cite{docker} or singularity \cite{singularity}, but only works as long as the container
image is preserved. Another downside of container solutions is that they usually ship with numerous system libraries which need to be updated or reproduced in case
of an image rebuild.

The later reproduction of scientific results that were obtained via a computer program requires
not only to follow the computational procedure, but also the exact same version of
the program. This can only be guaranteed if one is able to reproduce the executable of the
program.
Package managers, which aim at creating reproducible build environments, are Nix \cite{eelco} and Guix \cite{guix,guix2}.
These package managers are built around functional languages, which are used to create build recipes.
Such a build recipe is represented by a functional expression with inputs (e.g., other packages) and outputs
(a path with the build product), which are tracked with cryptographic hashes.
This approach to package management allows to uniquely identify
a particular build of a program and provides all prerequisites
to accurately reproduce the build at a later point in time.
Exact binary reproducibility is difficult to achieve with traditional package managers,
but can be achieved with the mechanisms provided by Nix.

In this paper, we show how the Nix package manager can be used to manage software in a sustainable and reproducible way. Our approach focuses mainly
on quantum chemistry software packages, but can be applied to any software.
We will introduce the \emph{NixOS-QChem} overlay, which is an add-on to the \emph{nixpkgs} collection for integrating quantum chemistry software into \emph{nixpkgs} environment and providing
optimized versions of the packages. The \emph{nixpkgs} collection is a good starting point since
it already provides $\approx$60,000 packages.
The add-on repository provides more build recipes for open source and proprietary software packages.

The paper is organized as follows. In sec. \ref{sec:intro} we give a general overview of the
Nix package manager and its features. In sec. \ref{sec:overlay} we describe the approach
to integrate quantum chemistry software packages along with a list of packaged software
(sec. \ref{sec:software}), followed by a set of examples in sec. \ref{sec:examples}.
In sec. \ref{sec:Comparison} we discuss how our approach compares to other packaging methods.

\section{Overview over the Nix package manager}
\label{sec:intro}
The Nix package manager \cite{eelco} is built around the Nix functional language \cite{nix} and
a set of packages can be represented by a set of functions, which eventually evaluate to file system paths.
A build recipe is called a Nix expression in the Nix terminology. Such Nix expressions are functions that evaluate to derivations, which uniquely describe the build process.
Nix expressions can have
one or more inputs and one or more outputs. The inputs are commonly other packages that provide dependencies, such as libraries. The output is a path to a final build product, such as the binary files.
The name of the output path is derived from a hash function over the derivation itself and all its inputs, which creates a unique path name.

The Nix package manager stores its packages (i.e., its output paths) under a fixed path in the file system,
\verb|/nix/store|, which is simply called the Nix store. Packages are not allowed to refer to dependencies outside the Nix store, thus avoiding dependencies with the system software.
All builds that are stored in the Nix store are immutable and can not be changed after a build
is completed, thus guaranteeing full stability of a given package.

The dependencies between Nix store paths are tracked in a local database for proper
handling by the Nix package manager.
Every package (and every variation of it) is stored under a unique path name.
As a result, many versions or variations of the same package can be installed
in parallel in the Nix store without interfering with the operating system's packages
or interference between Nix store paths.
Nix provides several mechanisms that allow for a user-friendly composition of the paths
in the Nix store into an environment for individual users.
Only a selected set of Nix store paths is projected into a user environment.
Packages are either built on demand or downloaded from a binary cache, thus making the manual installation of a package unnecessary.
No installation procedure by an administrator is thus necessary. This enables the end-user to use packaged software as well as to create their own builds.

Nix is, by design, a package manager which builds packages from source. If the output of
a build is not available in the local Nix store or in a remote binary cache, it will
be built from source. This means that if a dependency of a package changes, the package will
be rebuilt and potential problems with the update are either avoided or uncovered.

Note that the Nix approach differs substantially from the use of containers, which only statically
bundle dependencies but provide no further mechanism to update, rebuild and maintain
the contents of a container.

The second important component, besides Nix, is the \emph{nixpkgs} package collection \cite{nixpkgs}, which provides over 64,000 packages in the form of Nix expressions \cite{repology}
(including several quantum chemistry programs) and provides the basis for our work.
The packages provided by the \emph{nixpkgs} package collection are also available
in form of binaries through a binary cache and thus do not need to be build from source
by the end-user.
This package set can be extended by the user with the help of overlays that allows us
to add, modify, or replace packages.
We will introduce \emph{NixOS-QChem} overlay in sec \ref{sec:overlay}, and show how it is used to add and optimize packages.

\subsection{Nix in an HPC environment}
In this section, we address the challenges that arise in a high performance computer cluster
environment and how they can be addressed with Nix.

Environment modules \cite{modules} are an approach commonly used by super computing centers and on
scientific computer clusters to create on-demand environments for users. A module sets
environment variables pointing to the requested package paths upon a \verb|module load <package>| call. However, a significant shortcoming of this approach is that it does not track dependencies
between modules or any dependencies with system libraries.
As a consequence, even a minor system update, addressing only security updates,
may silently break installed packages or software compiled by a user.

The Nix package manager explicitly addresses these shortcomings. Users can choose to
build their own packages with Nix or use a centrally provided package set.
These user builds are independent of the operating system's software or centrally
installed packages.
The immutability of the Nix store guarantees that a dependency on an
existing package (Nix store path) can never be quietly replaced. This allows users to pin
a package and all its dependencies to a fixed version, providing stability and reproducibility
of the binaries.
When a dependent derivation is replaced or upgraded, it forces the rebuild of all
derivations which depend on it, ensuring a valid build.
The Nix package manager also provides an easy path for portability: packages can be transferred between different compute cluster systems (assuming that the Nix package manager is installed on both systems) either as binary or by automatically rebuilding the required packages
from their source.
Nix store paths and their dependencies can be transferred between machines by means of a custom copy command (\verb/nix copy --[to|from] <machine> <store path>/).

Note that a proper installation of the Nix package manager requires administrator rights and thus has to be carried out by a system administrator. The Gricad facility in Grenoble has
demonstrated \cite{Bzeznik17} how the Nix package manager can be used on a computer cluster with a
shared Nix store. Nix has been used for example at CERN to distribute software for LHCb \cite{CERN}.

\section{Structure of the Overlay and \emph{nixpkgs}}
\label{sec:overlay}
To customize \emph{nixpkgs} for use with quantum chemistry software packages, we make use of the
overlay mechanism, which allows us to extend and modify the package set provided by \emph{nixpkgs}.
Note that many scientific libraries and some quantum chemistry packages are already
packaged in \emph{nixpkgs}. These packages can be used directly after the installation of the
Nix package manager.
Our overlay, \emph{NixOS-QChem} \cite{nix-qchem}, is thus tightly coupled to \emph{nixpkgs}.
The overlay serves multiple purposes: it selects quantum chemistry related software packages for optimization
and adds additional quantum chemistry software packages that are not available in \emph{nixpkgs}.
The overlay also serves as an incubator for new packages that need to be matured first with
respect to its integration into the \emph{nixpkgs} environment. This includes packages that have
non-standard build systems and are thus more difficult to integrate.
The aim is to integrate a useful variety of  quantum chemistry packages into \emph{nixpkgs}
collection and to maintain a high code quality of the corresponding \emph{nixpkgs} guidelines.

\emph{NixOS-QChem} focuses on providing packages for the x86-64 CPU architecture on the Linux platform, as this is currently the most common architecture for scientific high-performance computing \cite{top500CPU}.
The overlay also provides optional performance optimizations, which make use of modern x86-64 processors, that are not provided by \emph{nixpkgs} itself due to compatibility reasons.
The optimizations allow for setting custom compiler flags and to automatically select optimization flags provided by individual packages.

All packages provided through the overlay are projected into a package subset
(name prefix: \verb|qchem|), which allows to also optimize basic libraries, such as the fftw
library \cite{fftw}, without causing the rebuild of non-scientific software packages.
Open source packages can be downloaded automatically from the internet, but
proprietary packages which require a license need to be provided by the user.
For these cases, the overlay also provides a mechanism which allows for downloading from a custom, internal location.
As a result, \emph{NixOS-QChem} can provide Nix expressions (and builds) for commercial packages - such as Turbomole, Molpro and others - as well as packages that require user registration - such as CFour, MRCC, and ORCA \cite{cfour, molpro, orca}.
While such packages often exclude the hurdles of compilation, packaging them enhances their composability.

Composing different major software packages in a single, coherent environment often proves difficult:
the problems range from different providers of MPI and BLAS/LAPACK implementations for different codes and name conflicts in a global \verb|$PATH| (e.g. \verb|libblas.so|, \verb|mpiexec|, ...), over different version constraints of the same dependencies (e.g. different version constraints of \verb|numpy| \cite{numpy} in different python packages, that cannot be fulfilled simultaneously).
In those cases correct behaviour can become dependent on detailed choices, such as in which order different environment modules are loaded.

For selected set of packages, we have implemented automated tests in the overlay, which ensure that the basic functionality of a package is still given after an update or a rebuild.
These tests are less comprehensive than the test suites provided by individual quantum chemistry packages, but aim at uncovering potential problems in connection with dependencies that have been observed during the integration.

The \emph{nixpkgs} repository provides a simple mechanism to switch between libraries, either on a
per-package basis or globally, for the whole package set. One example is
the message passing interface system (MPI) \cite{mpi}, which is provided by different implementations \cite{nix-mpi}. The default implementation is OpenMPI \cite{openmpi}
but it can be readily replaced by the overlay mechanism.
The following example shows the Nix code for an overlay that replaces OpenMPI
globally with MVAPICH \cite{mvapich} and builds the CP2K \cite{cp2k} package explicitly with MPICH \cite{mpich2}:
\begin{lstlisting}
self: super: {
  mpi = super.mvapich;
  cp2k = super.cp2k.override { mpi = self.mpich; };
}
\end{lstlisting}

Linear algebra libraries, such as BLAS and LAPACK can be replaced in a similar way.
Nixpkgs has a wrapper for BLAS and LAPACK \cite{blas, lapack99}, which provides custom libraries through the standard interface. The default implementation is OpenBLAS \cite{openblas}, but Intel's MKL \cite{mkl} or AMD's blis/libFlame \cite{blisflame}
are also available.
The following example demonstrates how an overlay can be used to replace BLAS and LAPACK with MKL:
\begin{lstlisting}
self: super: {
  blas = super.blas.override {
    blasProvider = self.mkl;
  };
  lapack = super.lapack.override {
    lapackProvider = self.mkl;
  };
}
\end{lstlisting}

The Nix code in the \emph{NixOS-QChem} overlay \cite{nix-qchem} is structured as follows:
\begin{itemize}
\item \verb|default.nix|: the base of the overlay
\item \verb|cfg.nix|: defines all configuration options for the overlay
\item \verb|nixpkgs-opt.nix|: defines all packages from the \emph{nixpkgs} collections
that are projected into the \verb|qchem| subset and are subject to processor dependent optimisations.
\item \verb|tests/|: folder with tests for various packages.
\item \verb|examples/|: folder with examples showing different configuration scenarios.
\item \verb|pkgs/|: contains sub folders with Nix expressions for additional packages.
\item \verb|install.sh|: installs Nix, nixpkgs and the NixOS-QChem overlay.
\end{itemize}

\subsection{List of Packaged Software}
\label{sec:software}
In combination, \emph{nixpkgs} and \emph{NixOS-QChem} provide a set of packages for quantum chemistry, molecular dynamics, and quantum dynamics (see table \ref{tab:PackageList} for a subset of packages) that can be used directly in a production
environment.
The Nix expressions describe how to build software from source, and the builds are executed on demand.
The package set is not restricted to free software and includes also Nix expressions
for proprietary packages.
Many packages profit from this integrated packaging, and composing coherent runtime environments is simplified.
Noteworthy examples for improved composability are the Pysisyphus optimiser \cite{pysisyphus}, which wraps Turbomole, ORCA, and Psi4 among others, or the polarizable LICHEM QM/MM implementation, which relies on the Tinker MM engine and the Gaussian, NWChem, and Psi4 quantum chemistry codes.
The SHARC surface hopping code, which depends on electronic structure codes,
can be used conveniently from \emph{NixOS-QChem}. SHARC requires deprecated Python2 as well as free and proprietary quantum chemistry engines (BAGEL, Molcas, ORCA, Turbomole, Gaussian, Molpro).
Using Nix an isolated environment with Python2 dependencies is available, quantum chemistry engines are directly provided to the SHARC scripts, and proper environment variables are set automatically, thus avoiding the error prone and difficult installation of multiple large quantum chemistry codes by the user.
To the best of the authors knowledge, the support for this variety of free and proprietary packages makes \emph{NixOS-QChem} unique among such packaging efforts.
With the DebiChem project of the Debian GNU/Linux distribution \cite{debichem}, another major packaging effort for chemical software exists.
While the DebiChem team often provides valuable knowledge and patches, the architecture and philosophy of Debian packaging prevents clean isolation and tight integration between packages.
Note that traditional package managers such as the Debian package manager are meant to be operated by system administrators and thus provide no straight forward way for end-user installations on a shared computer cluster.

\begin{table}
  \caption{List of selected quantum chemistry packages and utilities provided by the overlay.}
  \label{tab:PackageList}
  \begin{tabular}{lll}
    \toprule
    Package                 & Attribute          & Reference\\
    \midrule
    avogadro2-1.95.1        & qchem.avogadro2    &\cite{avogadro2} \\
    bagel-1.2.2             & qchem.bagel        &\cite{bagel1,bagel2}\\
    cfour-2.1               & qchem.cfour        &\cite{cfour} \\
    cp2k-8.2.0              & qchem.cp2k         &\cite{cp2k} \\
    crest-2.11.1            & qchem.crest        &\cite{crest} \\
    dalton-2020.0           & qchem.dalton       &\cite{dalton} \\
    ergoscf-3.8             & qchem.ergoscf      &\cite{ergoscf}\\
    gpaw-21.6.0             & qchem.gpaw         &\cite{gpaw} \\
    gromacs-2021.4          & qchem.gromacs      &\cite{gromacs} \\
    nwchem-7.0.2            & qchem.nwchem       &\cite{nwchem} \\
    molden-6.3              & qchem.molden       &\cite{molden} \\
    molpro-2021.2.0         & qchem.molpro       &\cite{molpro} \\
    mrcc-2020.02.22         & qchem.mrcc         &\cite{mrcc} \\
    octopus-11.2            & qchem.octopus      &\cite{octopus} \\
    openmolcas-21.10        & qchem.molcas       &\cite{openmolcas} \\
    orca-5.0.1              & qchem.orca         &\cite{orca} \\
    pcmsolver-1.3.0         & qchem.pcmsolver    &\cite{pcmsolver} \\
    psi4-1.4.1              & qchem.psi4         &\cite{psi4} \\
    pyscf-2.0.1             & qchem.python3.pkgs.pyscf &\cite{pyscf} \\
    pysisyphus-0.7.2        & qchem.pysisyphus   &\cite{pysisyphus} \\
    quantum-espresso-6.6    & qchem.quantum-espresso &\cite{quantumespresso} \\
    sharc-2.1.1             & qchem.sharc        &\cite{sharc} \\
    siesta-4.1-b3           & qchem.siesta       &\cite{siesta} \\
    tinker-8.8.3            & qchem.tinker       &\cite{tinker} \\
    turbomole-7.5.1         & qchem.turbomole    &\cite{turbomole} \\
    vmd-1.9.3               & qchem.vmd          &\cite{vmd} \\
    xtb-6.4.1               & qchem.xtb          &\cite{xtb} \\
    \bottomrule
  \end{tabular}
\end{table}

\subsection{Usage examples}
\label{sec:examples}
We will outline the basic installation procedure of the Nix package manager and
the overlay for a simple setup on a single machine. For the setup on a compute cluster, we refer to the setup of the Gricad team \cite{Bzeznik17} for a shared Nix store.
A multiuser installation of Nix can be obtained with the following commands:
\begin{lstlisting}
# To be executed by an admin
# Multi-user installation of Nix.
# Will request root privileges for the inital setup.
sh $> curl -L https://nixos.org/nix/install |
        sh -s -- --daemon
\end{lstlisting}
These commands will install Nix in multiuser mode; the Nix daemon will listen for evaluation requests from the Nix commands and execute builds or download the store
paths from a binary cache.

The packages in \emph{NixOS-QChem} can be accessed with different methods.
We will discuss two main methods here:
as a direct system-wide or user-installed overlay to the \emph{nixpkgs} channel and
explicit use as a project-based package source. The first method allows for
a direct use of the latest package versions, while the second method allows to fix the
version on a per-project basis.
Other options to access \emph{NixOS-QChem} overlay packages, which we will not discuss here further in detail, are the Nix user repositories (NUR) \cite{nur}, the experimental Nix flakes feature, or a customized Nix channel.
None of the above variants is mutual exclusive and each one can be useful for different scenarios.
NUR and the channel mechanism provide convenient automatic updates, while
flakes provide hermetic expressions, that are not influenced by the runtime environment, and the implicit overlay yields a convenient compostion of the package set.

The \emph{NixOS-QChem} overlay can be used as an implicit overlay by placing the repository in a directory recognized by \emph{nixpkgs}:
\begin{lstlisting}
sh $> mkdir -p ~/.config/nixpkgs/overlays
sh $> git clone \
      https://github.com/markuskowa/NixOS-QChem.git \
      ~/.config/nixpkgs/overlays/qchem
\end{lstlisting}
Packages from the overlay are then available for use via Nix commands, e.g. \verb|nix-shell -p qchem.xtb|.
Updates to the overlay happen explicitly by calling \verb|git pull|.
The behaviour of \emph{nixpkgs} and \emph{NixOS-QChem} can be controlled by settings in \verb|~/.config/nixpkgs/config.nix|, which allows to enable the build of proprietary packages and apply CPU related tuning options.
The Nix code of a configuration that enables AVX2 performance tuning for CPUs
from the Haswell generation onwards, and allows using proprietary packages,
is given by the following example:
\begin{lstlisting}
{
  config = {
    # Attempt build of packages
    # with non open source licences
    allowUnfree = true;
    qchem-config = {
      # Enable AVX2 CPU optmisations
      # (Haswell CPU target).
      optAVX = true;
      # Molpro license token if available
      licMolpro = null;
    };
  }
}
\end{lstlisting}

Nix can serve different use cases for computational tasks:
installing a package in the user's environment, launching an isolated shell, interactive use of a program, or the noninteractive execution of programs in a resource manager like SLURM.
To exemplify some common use cases, we will refer to illustrative examples in the following.

\paragraph{Interactive Program Usage (Turbomole)}:
    Turbomole uses a set of interactive programs, such as \verb|define| and \verb|eiger|, to create input files and analyse output files.
    Furthermore, Turbomole requires environment variables such as \verb|$TURBODIR| and \verb|$PARA_ARCH| to be set.
    An interactive \verb|nix-shell| makes the Turbomole package available and reduces the required user input, by wrapping Turbomole with appropriate environment variables and settings:
    \begin{lstlisting}
# starts a interative nix-shell with Turbomole
sh $> nix-shell -p qchem.turbomole

# Turbomole commands can directly be used
# normal interaction with define
# e.g. set up a RI-ADC(2) calculation
nix-shell $> define
# ground state calculation
nix-shell $> ridft -smpcpus 4
# excited state calculation
nix-shell $> ricc2 -smpcpus 4
# interactive overview of results
nix-shell $> eiger
# will drop back to normal bash
nix-shell $> exit
\end{lstlisting}

  \paragraph{Non-Interactive Calculation (Molcas)}
    A non-interactive, Molcas calculation with OMP parallelism can directly be executed from a \verb|nix-shell|.
    The PyMolcas driver requires specific Python packages, such as \verb|six|, to be installed.
    Instead of globally installing Python dependencies, the Nix expression wraps the python scripts in an isolated Python runtime environment and can be used directly:
    \begin{lstlisting}
sh $> nix-shell \
        -p qchem.molcas \
        --run "OMP_NUM_THREADS=4
	             pymolcas molcas.inp"
\end{lstlisting}

  \paragraph{Interactive Python Session (MEEP)}
    Some scientific Python packages may be used interactively within an interpreter, e.g. to experiment with different settings.
    Packages such as MEEP \cite{meep}, that provide a Python API around a C/C++ code, are often difficult to install; they are not available from PyPi and require both Python and C/C++ tooling.
    MEEP can be used interactively from Python within a \verb|nix-shell|:
    \begin{lstlisting}
sh $> nix-shell \
          -p python3 python3.pkgs.numpy \
             qchem.python3.pkgs.meep \
          --run "python3"
python3 $> import numpy as np
python3 $> import meep as mp
python3 $> # ...
\end{lstlisting}

  \paragraph{Project-Based Calculation Environment with Fixed Versions}
    Computational environments, that are associated with a specific project,
    can strongly benefit from fixing all package versions in a
    custom environment.
    Projects can use different versions or variations of programs without interfering
    with a system level package set.
    Such a computational environment can in principle be defined in a single Nix file
    and thus be easily shared between coworkers.
    Fixing all program versions in such an environment also allows reproducing its
    results at a later point in time.
    Such an environment can be described by a \verb|shell.nix| file, which defines an environment for a \verb|nix-shell|.
    To achieve reproducibility, the versions of \emph{nixpkgs} and \emph{NixOS-QChem} must be fixed:
    \begin{lstlisting}
let
 # Reproducible, pinned import of the
 # NixOS-QChem overlay function
 gh = "https://github.com";
 qchemOvl = import (builtins.fetchGit {
  url = "${gh}/markuskowa/NixOS-QChem.git";
  name = "NixOS-QChem_2021-09-25";
  rev = "9604e9b7f8d6ea68f07d621e1f70a9ebf857efa0";
  ref = "master";
 });

 nixpkgs = import (builtins.fetchGit {
  url = "${gh}/NixOS/nixpkgs.git";
  name = "nixpkgs_2021-09-25";
  rev = "a3a23d9599b0a82e333ad91db2cdc479313ce154";
  ref = "nixpkgs-unstable";
 });

 pkgs = nixpkgs {
   overlays = [ qchemOvl ];
   config = { ... };
 };

in with pkgs; mkShell { ... }
\end{lstlisting}
    Here, the \verb|fetchGit| function is used to access a specific version of the overlay, and the \emph{nixpkgs} package set  (fixed by the respective \verb|rev| statements).
    Alternatively, the Niv tool \cite{niv} provides a convenient command line interface to automate the version fixing and update processes.
    The overlay and configuration settings are applied explicitly in the \verb|shell.nix| file.
    For the rather verbose, full example of the \verb|shell.nix| file and the usage of Niv, we refer to \cite{Example-ProjectPin}.
    The \verb|shell.nix| file can either be referenced implicitly by executing \verb|nix-shell| in the same directory, or explicitly by \verb|nix-shell /path/to/shell.nix|.

  \paragraph{Reproducible Jupyter Notebooks}
    Jupyter notebooks \cite{jupyterNotebooks} are commonly used tools for experimentation with codes and methods, the development of scientific ideas,
    as well as for visualization of data.
    However, distributing Jupyter notebooks can be difficult,
    since the environment and all dependencies, such as Python packages, also need to be reproduced.
    Like in the previous example, \verb|nix-shell| can be used to make Jupyter notebooks reproducible.
    Using version fixing, as in the example above, a Jupyter environment for GPAW simulations can be formulated in a \verb|shell.nix| file:
    \begin{lstlisting}
let
  qchemOvl = ...
  pkgs = ...
  pythonWithPackages =
    pkgs.qchem.python3.withPackages
    (p: with p; [
      numpy
      jupyterlab
      ipympl
      gpaw
    ]);
in with pkgs; mkShell {
  buildInputs = [ pythonWithPackages ];
  shellHook = "jupyter-lab";
}
\end{lstlisting}
    Executing \verb|nix-shell| will then directly open the Jupyter-Lab interface in the browser and allow using packages such as GPAW, along with Python and all the
    necessary Python packages.
    The complete examples can be found in Ref. \cite{Example-ProjectPin}.

  \paragraph{Self-Contained Programs and Shell Scripts}
    The \verb|nix-shell| command can be used as the shebang line of scripts.
    This allows to write small, reproducible, self-contained scripts and programs, or to write scripts in the scope of a project-associated \verb|shell.nix| file.
    The following example shows a self-contained Python script for data visualization:
    \begin{lstlisting}
#! /usr/bin/env nix-shell
#! nix-shell -i python3
#! nix-shell -p python3Packages.numpy
#! nix-shell -p python3Packages.matplotlib

import numpy as np
import matplotlib.pyplot as plt

xs = np.linspace(-2, 2, num=100)
plt.plot(xs, np.exp(-xs**2))
plt.show()
\end{lstlisting}
    Note, that here we use the latest versions of numpy and matplotlib as they are provided directly by \emph{nixpkgs}.

    This mechanism can also be used to write SLURM (or other resource management system) scripts for working in a computer cluster environment. Such batch scripts can either
    pull in packages via \verb|nix-shell|'s \verb|-p| option or can be combined
    with a project-associated \verb|shell.nix| file
    \begin{lstlisting}
#! /usr/bin/env nix-shell
#! nix-shell /path/to/project/shell.nix -i bash

#SBATCH --ntasks=360
#SBATCH --ntasks-per-node=36
#SBATCH --nodes=10
#SBATCH --mem=0
#SBATCH --partition=s_standard

mpiexec \
  -np $SLURM_NTASKS \
  --map-by ppr:$SLURM_TASKS_PER_NODE:node \
  nwchem input.nw > output.log
\end{lstlisting}
    Reusing the environment from the project's \verb|shell.nix| ensures to have exactly the same computational environment on the compute nodes, the front-end node, or
    the user's local work station.
    This eliminates error-prone \verb|module load| operations, and ensures the independence of potentially different system libraries between nodes. These scripts can also be
    easily transferred between Nix enabled computing centers.

\section{Comparison with other Solutions}
\label{sec:Comparison}
\begin{table}
 \caption{Capabilities/characteristics of different software management options. (\yes) and (\no) indicate severe restrictions respective only very partial capabilities or workarounds.}
 \label{tab:Comparison}
 \begin{tabular}{l P{1.5em} P{1.5em} P{1.5em} P{1.5em} P{1.5em}}
    \toprule
                    & \rotatebox{90}{Nix} & \rotatebox{90}{Classical} & \rotatebox{90}{Spack} & \rotatebox{90}{Singularity} & \rotatebox{90}{Modules}   \\
    \midrule
    Mult. Versions  & \yes & \no      & \yes   & \yes        & \yes    \\
    Transferable    & \yes & \no      & (\yes) & \yes        & \no     \\
    Reproducible    & \yes & (\yes)   & \no    & (\yes)      & \no     \\
    Composable      & \yes & (\yes)   & (\yes) & \no         & (\no)   \\
    Customizable    & \yes & \no      & \yes   & \yes        & \yes    \\
    Multi-User      & \yes & \no      & \yes   & \yes        & \yes    \\
    \bottomrule
 \end{tabular}
\end{table}
Several approaches for managing software environments in high-performance computing have emerged, each offering specific advantages or disadvantages, inherited by their fundamental design.
In the following, we will compare the Nix packaging approach to other common approaches for managing software environments in high-performance computing.
The first category are distribution based package managers, such as the Debian package manager, Pacman for Arch Linux, and the Redhat package manager. Their main purpose is to provide software packages for a
system-wide installation.
A second approach is Spack, a solution explicitly designed with common HPC needs in mind.
Spack adopts concepts, such as dependency management, from distribution package managers, but allows users to manage their software and bring specific versions of programs into scope on demand.
Third are environment modules, which are commonly used on cluster systems to provide
customized packages. This solution does not provide software packages itself, but rather defines environment variables for use with locally installed packages.
Fourth are container solutions such as Docker and Singularity, which encapsulates an
application along with its dependencies in a whole operating system image. Singularity
has been specifically developed for HPC environments.
A fifth category are application level container formats such as Flatpack, Snap and AppImage, which were not adopted for use in HPC environments, and we will thus not discuss further here.

\paragraph{Dependency Handling}
Package managers, which follow a more traditional approach, such as the Debian package manager
and Spack, track dependencies and allow for clear dependency management. However, dependencies
may be replaced quietly without the rebuild of the package.
Environment modules may declare weak dependencies between each other but are not strictly required. Singularity itself is only a container
service, which does not define how software is build inside a container. It is thus up to the
user to choose an appropriate dependency tracking mechanism.
Nix has a strict dependency model, which is tracked with cryptographic hashes. Even a minor
change in a dependency forces a rebuild of a specific package.

\paragraph{Multiple Package Versions}
The ability to provide multiple versions of the same package is essential in a multiuser compute cluster environment.
Common examples include linear algebra libraries, different MPI implementations,
and different versions of compilers.
Traditional package managers, such as the Debian package manager, usually do not offer this capability,
as they install packages in a global name space.
Debian offers the \emph{update-alternatives} mechanism to switch, e.g., between different implementations of BLAS, but the behavior is global for all users of a given computer system.
Environment modules allow for parallel installation of identically named binaries and libraries
by using separate directory trees and composing an environment at run time.
Spack similarly allows loading packages in the namespace of a runtime environment, partially solving the problem with conflicts by stricter requirements on dependency and conflict declarations.
Singularity stores each application in an image file, thus avoiding collisions between different
images. Nix creates a separate store path for each build, strictly isolating different variants from each other. Runtime environments are then created on demand by using \verb|nix-shell|.

\paragraph{Transferability Between Systems}
The ability to share a computational environment enables other scientists to reproduce and validate scientific results.
Traditional package managers allow for an installation of identical packages on two different
machines only if they use the same Linux distribution with exactly the same version.
Environment modules can only be shared between different systems if they provide an identical installation.
Spack setups are slightly more distribution-agnostic: they assume common prerequisites in the underlying operating system, but can otherwise be self-contained.
However, as Spack does not use sandboxing techniques, an unknown amount of system-specific dependencies may leak in from the operating system.
In contrast, Nix packages and Singularity containers are fully transferable.
Singularity containers provide dependency bundling and are indepenendent of operating system libraries and are thus transferable. Nix ensures transferability by using unique store paths with a strict
set of dependencies, which can be transferred between machines.

\paragraph{Reproducibility}
Transferability and reproducibility are closely related. Here we are concerned with
the question if a program can be reproduced from scratch. This requires to rebuild
the program with exactly the same dependencies in the exact same build environment.
Environment modules and Spack do not provide any mechanism to achieve reproducibility.
Software installed via the distribution's package manager may be reproduced if the exact
state of all dependencies is recorded.
Singularity containers provide no specific mechanism to achieve reproducibility and it is
the responsibility of the user to define a procedure that leads to a reproducible result.
However, archiving a particular image may avoid the necessity to reproduce the build.
Nix ensures the reproducibility of the build environment by means of cryptographic hashes
on dependencies and sandboxing of the build procedure (see example "Project-Based Calculation Environment with Fixed Versions Computational environments").
Note, that \emph{nixpkgs} can also be used to build Singularity containers,
which are reproducible at the build stage.

\paragraph{Multi-User installation}
Spack, Singularity, and environment modules, as dedicated solutions for cluster computers, offer the option to provide software both centrally (e.g. a minimum set of basic tools and libraries from the local computer cluster environment), as well as on a per-user basis (customized or domain-specific software).
Neither approach takes advantage of shared common dependencies, which are identical between
packages. This can lead to increased storage requirements.
Nix packages are handled by a central daemon that controls access to the system-wide Nix store and executes builds on demand. This makes Nix an inherently multiuser solution and allows
every user to build of software, while Nix is intrinsically aware of shared dependencies.
Customized Nix packages can be made available by system administrators or can be created by
the user.

\section{Conclusion and Outlook}
The \emph{nixpkgs} set and the \emph{NixOS-QChem} overlay provide numerous scientific packages
and packages relevant for quantum chemistry. The presented solution makes these
programs easily available without complicated installation or
manual compilation procedures. Proprietary packages can also be made available without
explicit installation if the user has obtained a license and has the corresponding installation file.
The \emph{NixOS-QChem} overlay is configurable and allows for optimization depending on the used processor architectures. The option to build self-contained scripts and batch jobs
has proven itself highly useful in daily use. Its multiuser capability allows every
user to prepare customized and reproducible compute environments.
The presented examples demonstrate how to create
reproducible environments for electronic structure calculations as for scripted pre- and
post-processing tasks.

The presented approach is focused on applications for the theoretical chemistry community,
but the general principle is of broad applicability. We think that many scientific applications would benefit from the Nix approach. Reproducible environments are not
only useful for users of scientific software, but are also helpful during the
development of software.
Future developments of the \emph{NixOS-QChem} overlay
will aim at integrating more quantum chemistry software packages with Nix. The still experimental but under development and upcoming "Flakes" feature \cite{flakes} will simplify the creation of
reproducible environments with Nix even further.

We encourage users and developers of scientific software to contribute to \emph{NixOS-QChem}
and \emph{nixpkgs} as well as to report bugs.
It would be a great advantage if more computing facilities will adopt the approach and provide a Nix installation to allow for more reproducible compute environments.

\section{Acknowledgments}
The authors would like to thank the \emph{nixpkgs} community for providing the software infrastructure that made this work possible.
Phillip Seeber gratefully acknowledges the financial support provided by the German Research Foundation within the TRR CATALIGHT -- Projektnummer 364549901-TRR234 (project C5).


\end{document}